\documentclass[12pt]{article}
\usepackage{amssymb}
\begin{document}
\thispagestyle{empty}
\newcommand{\be}{\begin{equation}}
\newcommand{\ee}{\end{equation}}
\newcommand{\sect}[1]{\setcounter{equation}{0}\section{#1}}
\newcommand{\vs}[1]{\rule[- #1 mm]{0mm}{#1 mm}}
\newcommand{\hs}[1]{\hspace{#1mm}}
\newcommand{\mb}[1]{\hs{5}\mbox{#1}\hs{5}}
\newcommand{\bea}{\begin{eqnarray}}
\newcommand{\eea}{\end{eqnarray}}
\newcommand{\wt}[1]{\widetilde{#1}}
\newcommand{\ux}[1]{\underline{#1}}
\newcommand{\ov}[1]{\overline{#1}}
\newcommand{\sm}[2]{\frac{\mbox{\footnotesize #1}\vs{-2}}
           {\vs{-2}\mbox{\footnotesize #2}}}
\newcommand{\prt}{\partial}
\newcommand{\eps}{\epsilon}\newcommand{\p}[1]{(\ref{#1})}
\newcommand{\R}{\mbox{\rule{0.2mm}{2.8mm}\hspace{-1.5mm} R}}
\newcommand{\Z}{Z\hspace{-2mm}Z}
\newcommand{\cd}{{\cal D}}
\newcommand{\cg}{{\cal G}}
\newcommand{\ck}{{\cal K}}
\newcommand{\cw}{{\cal W}}
\newcommand{\vj}{\vec{J}}
\newcommand{\vl}{\vec{\lambda}}
\newcommand{\vz}{\vec{\sigma}}
\newcommand{\vt}{\vec{\tau}}
\newcommand{\poiss}{\stackrel{\otimes}{,}}
\newcommand{\tx}{\theta_{12}}
\newcommand{\tb}{\overline{\theta}_{12}}
\newcommand{\zw}{{1\over z_{12}}}
\newcommand{\sqp}{{(1 + i\sqrt{3})\over 2}}
\newcommand{\sqm}{{(1 - i\sqrt{3})\over 2}}
\newcommand{\NP}[1]{Nucl.\ Phys.\ {\bf #1}}
\newcommand{\PLB}[1]{Phys.\ Lett.\ {B \bf #1}}
\newcommand{\PLA}[1]{Phys.\ Lett.\ {A \bf #1}}
\newcommand{\NC}[1]{Nuovo Cimento {\bf #1}}
\newcommand{\CMP}[1]{Commun.\ Math.\ Phys.\ {\bf #1}}
\newcommand{\PR}[1]{Phys.\ Rev.\ {\bf #1}}
\newcommand{\PRL}[1]{Phys.\ Rev.\ Lett.\ {\bf #1}}
\newcommand{\MPL}[1]{Mod.\ Phys.\ Lett.\ {\bf #1}}
\newcommand{\BLMS}[1]{Bull.\ London Math.\ Soc.\ {\bf #1}}
\newcommand{\IJMP}[1]{Int.\ J.\ Mod.\ Phys.\ {\bf #1}}
\newcommand{\JMP}[1]{Jour.\ Math.\ Phys.\ {\bf #1}}
\newcommand{\LMP}[1]{Lett.\ Math.\ Phys.\ {\bf #1}}
\newpage
\setcounter{page}{0} \pagestyle{empty} \vs{12}
\begin{center}
{\Large {\bf Hermitian versus holomorphic complex and quaternionic generalized supersymmetries of the M-theory. A classification.}}\\ {\quad}\\

\vs{10} {\large Francesco Toppan} ~\\ \quad
\\
 {\large{\em CCP - CBPF}}\\{\em Rua Dr. Xavier Sigaud
150, cep 22290-180 Rio de Janeiro (RJ)}\\{\em Brazil}\\

\end{center}
{\quad}\\ \centerline{ {\bf Abstract}}

\vs{6}

Relying upon the division-algebra classification of Clifford algebras and spinors,
a classification of generalized supersymmetries (or, with a slight
abuse of language,``generalized supertranslations") is provided.
In each given space-time the maximal, saturated, generalized supersymmetry, 
compatible with the division-algebra constraint that can be consistently imposed 
on spinors and on superalgebra generators, is furnished.
Constraining the superalgebra generators in both the complex and the quaternionic cases gives rise to the 
two classes of constrained hermitian and holomorphic generalized supersymmetries.
In the complex case these two classes of generalized supersymmetries can be regarded as complementary.
The quaternionic holomorphic supersymmetry only exists in certain space-time dimensions and can admit
at most a single bosonic scalar central charge.\par
The results here presented pave the way for a better understanding of the various $M$ algebra-type
of structures which can be introduced in different space-time signatures
and in association with different division algebras, as well as their mutual relations. 
In a previous work, e.g., the introduction of a complex holomorphic
generalized supersymmetry was shown to be necessary in order to perform the analytic continuation of the standard $M$-theory
to the $11$-dimensional Euclidean space. As an application of the present results, it is shown that the above algebra
also admits a $12$-dimensional, Euclidean, $F$-algebra presentation.

\vs{6} \vfill \rightline{CBPF-NF-001/04}
{\em E-mail:}\quad {$toppan@cbpf.br$}\\
{\em Homepage:}\quad{$http://www.cbpf.br/\sim toppan$}
\pagestyle{plain}
\renewcommand{\thefootnote}{\arabic{footnote}}

\section{Introduction.}

The problem of classifying supersymmetries has recently regained
interest and found a lot of attention in the literature. 
We can cite, e.g., a series of papers where the notion of
``spin algebra" has been introduced and investigated \cite{fer}. An even more
updated reference concerns the classification of the so-defined ``polyvector super-Poincar\'e algebras"
\cite{acdp}.
\par
The reasons behind all this activity are clear. In the seventies
the H\L S scheme \cite{hls} was a cornerstone providing the supersymmetric
extension of the Coleman-Mandula no-go theorem. 
However, in the nineties, the generalized space-time supersymmetries going
beyond the H\L S scheme (and admitting, in particular, a bosonic sector of the
Poincar\'e or conformal superalgebra which could no longer be expressed as 
a tensor product $B_{geom}\oplus B_{int}$, where $B_{geom}$ describes
space-time Poincar\'e or conformal algebras, while the remaining generators
spanning $B_{int}$ are scalars) found widespread recognition \cite{{agit},{tow}}
in association with the dynamics of extended objects like branes (see \cite{{gs},{ste}}).
The eleven-dimensional $M$-algebra underlying the $M$-theory as a possible ``Theory 
Of Everything" (TOE), admitting $32$-real component spinors and maximal number ($=528$) of
saturated bosonic generators \cite{{agit},{tow}} falls into this class of generalized supersymmetries.
The physical motivations for dealing with and classifying generalized supersymmetries are therefore
quite obvious (later on, we will discuss some further evidence concerning the importance
of such a classification). The purely mathematical side as well presents very attracting features.
The ingredients that have to be used have been known by mathematicians since at least the sixties
(\cite{abs}, see also \cite{por} and, for quite a convenient presentation for physicists,
\cite{oku}). They include the division-algebra classification of Clifford algebras and fundamental
spinors.
It is quite rewarding that, by using these available tools, we can conveniently formulate
and solve
the problem of classifying generalized supersymmetries.
This is indeed the content of the present paper.\par
Before introducing the results here obtained, let me further elucidate the basic motivations.
In a series of previous works in collaboration with Lukierski \cite{lt1}-\cite{lt3}, we discussed the role
of division algebras (complex, quaternionic and octonionic) in the context of generalized supersymmetries.
The last work in particular deals with a problem not previously addressed in the literature,
namely how to perform the analytic continuation from the Minkowskian $11$-dimensional $M$-theory to
its Euclidean counterpart. We proved that this could be achieved by introducing the notion of
``$11$-dimensional Euclidean complex holomorphic supersymmetry". This is a constrained generalized
supersymmetry admitting a complex structure compatible with the one generated by the $32$ complex
Euclidean supercharges. In formulating an Euclidean version of the $M$-theory, whatever the latter
could possibly be, we should therefore naturally expect that the above Euclidean supersymmetry would play
a fundamental role. On the other hand, while eleven dimensions are singled out on a physical ground
through the TOE viewpoint previously recalled, on a purely algebraic level, no limitation 
concerning the dimensionality of the space-time is found
for the introduction of generalized supersymmetries algebras.
Indeed, specific examples of dynamical systems admitting generalized supersymmetries in different space-time
dimensions have been explicitly constructed in the literature (four-dimensional superparticle models with tensorial central charges have been constructed, e.g., in \cite{rs}, see also \cite{bl}).
It is therefore worth providing a purely mathematical classification of the generalized supersymmetries algebras
in each given space-time signature, as we will address here.
To be more specific, we classify here the most general supersymmetry algebra whose fermionic generators can be considered as the ``square roots" of the bosonic
generators, the latter being expanded as tensorial central charges. With a slight abuse of language, such structures
can also be referred to as ``generalized supertranslations" even if, strictly speaking, in some given space-time
signatures there are no translations at all. For instance, the $M$-algebra (\ref{Malg}) can be rexpressed in the $F$-viewpoint \cite{vafa}
in terms of $D=12$ dimensional (with $(10,2)$ signature) $32$-component Majorana-Weyl spinors. In this framework
the $528$ bosonic generators are split into $66$ rank-two antisymmetric tensors and $462$ self-dual rank six
antisymmetric tensors (\ref{Falg}). Therefore, in the $F$-viewpoint there are no vectors, which means no translations.\par
It is well-known that the Clifford algebra irreps \cite{oku} are put in correspondence with the ${\bf R},
{\bf C}, {\bf H}$ division algebras. An analogous scheme works for fundamental spinors (here and in the following,
fundamental spinors are defined to be the spinors admitting, in a given space-time, the maximal division algebra structure compatible with the minimal number of
{\em real} components). 
Both the $M$-algebra and the $F$-algebra presentations recalled above are based on real spinors, which makes life 
quite simple. On the other hand, the presence of both complex and quaternionic spinors allows intoducing division-algebra compatible extra-constraint on the available generalized supersymmetries. 
The reason for that lies in the fact that in these two extra cases one has at disposal the division-algebra principal conjugation (which simply coincides, for real numbers, with the identity operator)
to further play with. As a consequence, the two big classes of (complex or quaternionic) constrained hermitian versus
holomorphic generalized supersymmetries can be consistently introduced (see the end of Section {\bf 5} and the following sections).\par
It is of particular importance to determine the biggest (``saturated") generalized supersymmetry compatible with the given division-algebra structure and constraint. The complete classification is here presented in a series of tables.
We should recall that, in the $M$-algebra case, the Lorentz covariance requires that all combinations
of the $11$ vectors, $55$ rank-two and $462$ rank-five antisymmetric tensors are allowed as a consistent subalgebra while,
on the other hand, the $528=11+55+462$ bosonic components of the $M$-algebra provide the maximal admissible algebra.
A similar situation is faced in the complex and quaternionic cases. It is worth stressing that the hermitian/ holomorphic constraint mentioned before admits an algebraic interpretation and in this respect it can be 
regarded on a different plan w.r.t., let's say, 
the restriction of setting equal to zero all $462$ rank-five 
generators in (\ref{Malg}). The latter is an admissible, however hand-imposed, restriction, while the hermitian-versus-holomorphic constraint has a deeper division-algebra motivation. The recognition of this property could
become important in the future when investigating a certain class of dynamical systems (e.g., as a protecting mechanism towards the arising of extra anomalous terms).\par
For the sake of simplicity, in this work we are only concerned with ``generalized supertranslations", in the loose sense mentioned before. This means in particular that the bosonic generators are all abelian. The construction of,
e.g., Lorentz generators requires a bigger algebra than the ones here examined. One viable scheme to produce them consists in introducing a generalized superconformal algebra (which, in its turn, allows recovering a generalized superPoincar\'e algebra through an Inon\"u-Wigner type of contraction). Following \cite{lt1}, this can be easily achieved by taking two separate copies
of ``generalized supertranslations" and imposing the Jacobi identities on the whole set of generators to
fully determine the associated superconformal algebra. This scheme is rather straightforward and will not be addressed  here. It corresponds to technical developments which are left to future investigations.\par
The scheme of the paper is as follows: in Section {\bf 2} we review the basic properties of Clifford algebras and of their association with division algebras. An algorithm to explicitly construct the Clifford irreps in any $(s,t)$ space-time is introduced. Division algebras are quickly reviewed in Section {\bf 3}. In section {\bf 4} we discuss
the spinors-versus-Clifford algebras division-algebra character. Since our later construction is based on
matching the division-algebra properties of both spinors and Clifford algebras, we point out that, depending
on the given space-time, we can look at their maximal common division algebra character or, if necessary,
we can work either with reducible Clifford representations or with non-fundamental spinors. In Section {\bf 5}
generalized supersymmetries (real, complex and quaternionic) are introduced. The notion 
of consistently constrained hermitian and holomorphic (complex and quaternionic) generalized
supersymmetries is explained. These constrained supersymmetries are further discussed and classified in the 
following Sections {\bf 6} and {\bf 7}. A series of tables will be provided expressing their main properties, the total number of bosonic components, etc. The case of the quaternionic holomorphic generalized supersymmetry is particularly interesting. It will be proven that it can only exist in special dimensions of the space-time and that it can admit
at most a single (scalar) central extension.
In Section {\bf 8}, as a special application of our method, we prove that the Euclidean $M$-algebra of reference \cite{lt3}
finds a $12$-dimensional Euclidean $F$-algebra counterpart, which is a holomorphic supersymmetry based on
Weyl projected (in the generalized sense discussed in Section {\bf 2}) complex spinors. In the Conclusions we finally discuss some promising applications (ranging from higher-dimensional Chern-Simon supergravities, the theory of 
higher spins, some dynamical models based on supersymmetries with tensorial central charges, etc.)
of the formalism here introduced. 

\section{On Clifford algebras.}

The classification of generalized supersymmetries requires the preliminary classification
of Clifford algebras and spinors and of their association with division algebras.\par
To make this paper self-consistent, in this section we review the classification of the Clifford algebras
associated to the ${\bf R}, ${\bf C}, ${\bf H}$ associative
division algebras, following
\cite{oku} and \cite{crt1}. Some further material, which is useful for later purposes, will also be 
integrated.
\par The most general irreducible {\em real}
matrix representations of the Clifford algebra
\begin{eqnarray}
\Gamma^\mu\Gamma^\nu+\Gamma^\nu\Gamma^\mu &=& 2\eta^{\mu\nu},
\label{cliff}
\end{eqnarray}
with $\eta^{\mu\nu}$ being a diagonal matrix of $(p,q)$ signature
(i.e. $p$ positive, $+1$, and $q$ negative, $-1$, diagonal
entries)\footnote{Throughout this paper it will be understood that the positive eigenvalues are associated
with space-like directions, the negative ones with time-like directions.}
can be classified according to the property of the most
general $S$ matrix commuting with all the $\Gamma$'s ($\relax
[S,\Gamma^\mu ] =0$ for all $\mu$). If the most general $S$ is a
multiple of the identity, we get the normal (${\bf R}$) case.
Otherwise, $S$ can be the sum of two matrices, the second one
multiple of the square root of $-1$ (this is the almost complex,
${\bf C}$ case) or the linear combination of $4$ matrices closing
the quaternionic algebra (this is the ${\bf H}$ case).
According
to \cite{oku} the {\em real} irreducible representations are of
${\bf R}$, ${\bf C}$, ${\bf H}$ type, according to the following
table, whose entries represent the values $p-q~ mod~ 8$
 { {{\begin{eqnarray}&
\begin{tabular}{|c|c|c|}\hline
${\bf R} $&${\bf C}$ &${\bf H}$\\ \hline $0,2$&$$&$4,6$\\ \hline
$1$&$3,7$&$5$\\ \hline
\end{tabular}&\end{eqnarray}}} }

The real irreducible representation is always unique unless
$p-q~mod~8 = 1,5$. In these signatures two inequivalent real
representations are present, the second one recovered by flipping
the sign of all $\Gamma$'s ($\Gamma^\mu \mapsto - \Gamma^\mu$).

Let us denote as $C(p,q)$ the Clifford irreps corresponding to the $(p,q)$
signatures. The normal (${\bf R}$), almost complex (${\bf C}$) and quaternionic (${\bf H}$)
type of the corresponding Clifford irreps can also be understood as follows. While in the
${\bf R}$-case the matrices realizing the irrep have necessarily real entries,
in the ${\bf C}$-case matrices with complex entries can be used, while in the ${\bf H}$-case
the matrices can be realized with quaternionic entries.\par
Let us discuss the simplest examples. The ${\bf C}$-type $C(0,1)$ Clifford algebra
can be expressed either through the $2\times 2$ matrix with real-valued entries
$\left( \begin{array}{cc}
  0 & 1 \\-1 & 0
\end{array}\right)$ or through the imaginary number $i$.\par 
The ${\bf H}$-type Clifford algebra
$C(0,3)$, on the other hand, can be realized as follows:\\
{\em i)} with three $4\times 4$ matrices with real entries, given by the tensor products
$  \tau_A\otimes\tau_1$, $ \tau_A\otimes\tau_2$ and $ 
  {\bf 1}_2\otimes \tau_A$, where the matrices $\tau_A$, $\tau_1$ and $\tau_2$ furnish
  a real irrep of $C(2,1)$\par 
  ($\tau_A =\left(\begin{array}{cc}
  0 & 1 \\-1 & 0
\end{array}\right)$,
$\tau_1=\left(\begin{array}{cc}
  0 & 1 \\1 & 0
\end{array}\right)$,
$\tau_2=\left(\begin{array}{cc}
  1& 0\\0 & -1
\end{array}\right)$)
,\\
{\em ii)} with three $2\times 2$ complex-valued matrices given by
$\left( \begin{array}{cc}
  0 & 1 \\-1 & 0
\end{array}\right)$,
$\left( \begin{array}{cc}
  0 & i \\i & 0
\end{array}\right)$ and
$\left( \begin{array}{cc}
  i& 0 \\0 & i
\end{array}\right)$,
\\
{\em iii)} with the three imaginary quaternions $e_i$ (see for more details the section {\bf 3}).\par
The formulas at the items {\em i)} and {\em ii)} provide the real and complex representations, respectively,
for the imaginary quaternions. They can be straightforwardly extended to provide real and complex representations
for the ${\bf H}$-type Clifford algebras by substituting the quaternionic entries with the corresponding
representations (the quaternionic identity $1$ being replaced in the complex representation by the 
$2\times 2$ identity matrix ${\bf 1}_2$ and by the $4\times 4$ identity matrix ${\bf 1}_4$ in the real
representation).\par
It is worth noticing that in the given signatures $p-q~mod~8 = 0,4,6,7$,
without loss of generality, the $\Gamma^\mu$ matrices can be
chosen block-antidiagonal (generalized Weyl-type matrices), i.e.
of the form
\begin{eqnarray}
\Gamma^\mu &=&\left( \begin{array}{cc}
  0 & \sigma^\mu \\
  {\tilde\sigma}^\mu & 0
\end{array}\right)\label{weyl}
\end{eqnarray}
In these signatures it is therefore possible to introduce the
Weyl-projected spinors, whose number of components is half of the
size of the corresponding $\Gamma$-matrices\footnote{This notion of Weyl spinors,
which is convenient for our purposes, is
different from the one usually adopted in connection with {\em complex}-valued Clifford algebras
and
has been introduced in \cite{crt1}.}.\par
A very convenient presentation of the irreducible representations of Clifford algebras
with the help of an algorithm allowing to single out, in each arbitrary signature space-time,
a representative (up to, at most, the sign flipping $\Gamma^{\mu}\leftrightarrow -\Gamma^{\mu}$)
in each irreducible class of representations of Clifford's gamma matrices has been given in \cite{crt1}.
We recall and extend here the results presented in \cite{crt1}, making explicit the connection
between the maximal-Clifford algebras in the table (\ref{bigtable}) below and their division-algebra 
property. 
\par
The construction goes as follows.
At first one proves that starting from
a given $D$ spacetime-dimensional representation of Clifford's
Gamma matrices, one can recursively construct $D+2$ spacetime
dimensional Clifford  Gamma matrices with the help of two
recursive algorithms. Indeed, it is a simple exercise to verify
that if $\gamma_i$'s denotes the $d$-dimensional Gamma matrices of
a $D=p+q$ spacetime with $(p,q)$ signature (namely, providing a
representation for the $C(p,q)$ Clifford algebra) then
$2d$-dimensional $D+2$ Gamma matrices (denoted as $\Gamma_j$) of a
$D+2$ spacetime are produced according to either
\begin{eqnarray}
 \Gamma_j &\equiv& \left(
\begin{array}{cc}
  0& \gamma_i \\
  \gamma_i & 0
\end{array}\right), \quad \left( \begin{array}{cc}
  0 & {\bf 1}_d \\
  -{\bf 1}_d & 0
\end{array}\right),\quad \left( \begin{array}{cc}
  {\bf 1}_d & 0\\
  0 & -{\bf 1}_d
\end{array}\right)\nonumber\\
&&\nonumber\\ (p,q)&\mapsto&
 (p+1,q+1).\label{one}
\end{eqnarray}
or
\begin{eqnarray}
 \Gamma_j &\equiv& \left(
\begin{array}{cc}
  0& \gamma_i \\
  -\gamma_i & 0
\end{array}\right), \quad \left( \begin{array}{cc}
  0 & {\bf 1}_d \\
  {\bf 1}_d & 0
\end{array}\right),\quad \left( \begin{array}{cc}
  {\bf 1}_d & 0\\
  0 & -{\bf 1}_d
\end{array}\right)\nonumber\\
&&\nonumber\\ (p,q)&\mapsto&
 (q+2,p).\label{two}
\end{eqnarray}
It is immediate to notice that the three matrices $\tau_A$, $\tau_1$, $\tau_2$ 
introduced before and realizing the Clifford algebra $C(2,1)$ are obtained by applying
either (\ref{one}) or (\ref{two}) to the number $1$, i.e. the
one-dimensional realization of $C(1,0)$. 
\par
All Clifford algebras of ${\bf R}$-type are obtained by recursively applying the
algorithms (\ref{one}) and (\ref{two}) to the Clifford algebra
$C(1,0)$ ($\equiv 1$) and the Clifford algebras of the series
$C(0, 7+8m)$ (with $m$ non-negative integer), which must be
previously known. Similarly, all Clifford algebras of
${\bf H}$-type are obtained by recursively applying the algorithms to the Clifford algebras
$C(0, 3+8m)$, while the ${\bf C}$-type Clifford algebras are obtained by recursively
applying the algorithms to the Clifford algebras $C(0, 1+8m)$ and $C(0,5+8m)$.
 This is in accordance with the scheme
illustrated in the table below, taken from \cite{crt1}. Actually, the table
below is slightly modified w.r.t. the one in \cite{crt1}, since for our purposes here we
need to emphasize the complex character of the ${\bf C}$-type Clifford algebras.
We get  \par {~}\par {\em Table with the
maximal Clifford algebras (up to $d=256$).}

 {\tiny{
{\begin{eqnarray} &
\begin{tabular}{|c|cccccccccccccccccccc||}
\hline
   &1  &$\ast$& 2&$\ast$&  4&$\ast$& 8&$\ast$&16&$\ast$&32&$\ast$&64&
   $\ast$&128&$\ast$&256&$\ast$
      \\ \hline
      
 & &&&&&&&&&&&&&\\
 {\bf R} &$\underline{(1,0)}$ &$\Rightarrow$& $(2,1)$ &$\Rightarrow$&(3,2)
 &$\Rightarrow$&
  (4,3) &$\Rightarrow$&(5,4)& $\Rightarrow$ &(6,5)
&$\Rightarrow$&
  (7,6) &$\Rightarrow$&(8,7)& $\Rightarrow$ &(9,8)
  &$\Rightarrow$

  \\
  &&&&&&&&&&\\
 
 \hline
  &&&&&&&&&&\\

   & &&&&(1,2)&$\rightarrow $&(2,3)&$\rightarrow$&(3,4)
&$\rightarrow $&(4,5)&$\rightarrow$&(5,6)&$\rightarrow
$&(6,7)&$\rightarrow$&(7,8)&$\rightarrow$
   \\
  &&&&$\nearrow$&&&&&&&\\
  {\bf C}&&&{\underline{(0,1)}}&& &&&&&&\\
  &&&&$\searrow$&&&&&&&\\
  &&&&&&&&&&\\
  &&&&&(3,0)&$\rightarrow
  $&(4,1)&$\rightarrow$&(5,2)&$\rightarrow$
&(6,3)&$\rightarrow $&(7,4)&$\rightarrow$&(8,5)&$\rightarrow$&(9,6)& $\rightarrow$
  \\
   &&&&&&&&&&\\
   
   \hline
  &&&&&&&&&&\\

   &&&&&&&(1,4)&$\rightarrow $&(2,5)&$\rightarrow$&(3,6)
&$\rightarrow $&(4,7)&$\rightarrow$&(5,8)&$\rightarrow
$&(6,9)&$\rightarrow$
   \\
  &&&&&&$\nearrow$&&&&&\\
 {\bf H} &&&&&{\underline{(0,3)}}&& &&&&\\
  &&&&&&$\searrow$&&&&&\\
  &&&&&&&&&&\\
  &&&&&&&(5,0)&$\rightarrow
  $&(6,1)&$\rightarrow$&(7,2)&$\rightarrow$
&(8,3)&$\rightarrow $&(9,4)&$\rightarrow$&(10,5)&$\rightarrow$
  \\
   &&&&&&&&&&\\
   
   \hline
  &&&&&&&&&&\\

   &&&&&&&&&(1,6)&$\rightarrow $&(2,7)
&$\rightarrow $&(3,8)&$\rightarrow$&(4,9)&$\rightarrow
$&(5,10)&$\rightarrow$
   \\
  &&&&&&&&$\nearrow$&&&\\
  {\bf C}&&&&&&&{\underline{(0,5)}}&& &&\\
  &&&&&&&&$\searrow$&&&\\
  &&&&&&&&&&\\
  &&&&&&&&&(7,0)&$\rightarrow $&(8,1)
&$\rightarrow $&(9,2)&$\rightarrow$&(10,3)&$\rightarrow
$&(11,4)&$\rightarrow$
\\
&&&&&\\ 
\hline
&&&\\

   &&&&&&&&&(1,8)&$\rightarrow $&(2,9)
&$\rightarrow $&(3,10)&$\rightarrow$&(4,11)&$\rightarrow
$&(5,12)&$\rightarrow$
   \\
  &&&&&&&&$\nearrow$&&&\\
  {\bf R}/{\bf O}&&&&&&&{\underline{(0,7)}}&& &&\\
  &&&&&&&&$\searrow$&&&\\
  &&&&&&&&&&\\
  &&&&&&&&&(9,0)&$\rightarrow $&(10,1)
&$\rightarrow $&(11,2)&$\rightarrow$&(12,3)&$\rightarrow
$&(13,4)&$\rightarrow$
\\
&&&&&\\ 
\hline
&&&\\

  &&& &&&&&&&&&&(1,10)&$\rightarrow $&(2,11)
&$\rightarrow $&(3,12)&$\rightarrow$
   \\
  &&&&&&&&&&&&$\nearrow$&&&&&\\
 
  {\bf C}&&&&&&&&&&&{\underline{(0,9)}}&&&& &&\\
  &&&&&&&&&&&&$\searrow$&&&&&\\
  &&&&&&&&&&&&&&\\
  &&&&&&&&&&&&&(11,0)&$\rightarrow $&(12,1)
&$\rightarrow $&(13,2)&$\rightarrow$\\ &&&&&\\
\hline
 &&&\\

  &&&&& &&&&&&&&&&(1,12)&$\rightarrow $&(2,13)
&$\rightarrow $
   \\
  &&&&&&&&&&&&&&$\nearrow$&&&\\
 
  {\bf H}&&&&&&&&&&&&&{\underline{(0,11)}}&& &&\\
  &&&&&&&&&&&&&&$\searrow$&&&\\

  &&&&&&&&&&&&&&\\
  &&&&&&&&&&&&&&&(13,0)&$\rightarrow $&(14,1)
&$\rightarrow $\\ &&&&&\\

\hline  &&&\\
 &&&& &&&&& &&&&&&&&(1,14)&$\rightarrow $
   \\
  &&&&&&&&&&&&&&&&$\nearrow$&&&\\
  {\bf C}&&&&&&&&&&&&&&&{\underline{(0,13)}}&& &&\\
  &&&&&&&&&&&&&&&&$\searrow$&&&\\

  &&&&&&&&&&&&&&\\
  &&&&&&&&&&&&&&&&&(15,0)&$\rightarrow $\\

  &&&&&\\
  \hline &&&\\
 &&&& &&&&& &&&&&&&&(1,16)&$\rightarrow $
   \\
  &&&&&&&&&&&&&&&&$\nearrow$&&&\\
 {\bf R}/{\bf O} &&&&&&&&&&&&&&&{\underline{(0,15)}}&& &&\\
  &&&&&&&&&&&&&&&&$\searrow$&&&\\
  &&&&&&&&&&&&&&\\
  &&&&&&&&&&&&&&&&&(17,0)&$\rightarrow $\\
  \hline

\end{tabular}&\nonumber
\end{eqnarray}}}}
\begin{eqnarray}\label{bigtable}
&& \end{eqnarray} 

Concerning the above table some remarks are in order. The columns are labeled by
the matrix size $d$ (in real components) of the maximal 
Clifford algebras. Their signature is denoted by the
$(p,q)$ pairs. Furthermore, the underlined Clifford algebras in
the table can be named as ``primitive maximal Clifford algebras".
The remaining maximal Clifford algebras appearing in the table are
the ``maximal descendant Clifford algebras". They are obtained
from the primitive maximal Clifford algebras by iteratively
applying the two recursive algorithms (\ref{one}) and (\ref{two}).
Moreover, any non-maximal Clifford algebra is obtained from a
given maximal Clifford algebra by deleting a certain number of
Gamma matrices (this point has been fully explained in \cite{crt1} and will not be further
elaborated here).\par
The maximal Clifford algebras generated by the $C(0, 7+8m)$ series are associated to
both the real (${\bf R}$) and octonionic (${\bf O}$)
division algebras, since (\ref{cliff}), for the $(0, 7+8m)$-signature, can be realized either
associatively (in the normal, ${\bf R}$, case), or non-associatively through the octonions
(see \cite{crt1} and \cite{crt2}).\par
The primitive maximal Clifford algebras $C(0,3)$ and $C(0,7)$ can be explicitly realized through, respectively,
three $4\times 4$ matrices (as already recalled) and seven $8\times 8$ matrices given by
\begin{eqnarray}
C(0,3) &\equiv& \begin{array}{c}
  \tau_A\otimes\tau_1, \\
  \tau_A\otimes\tau_2, \\
  {\bf 1}_2\otimes \tau_A.
\end{array}
\end{eqnarray}
and
\begin{eqnarray}
C(0,7) &\equiv& \begin{array}{c}
  \tau_A\otimes\tau_1\otimes{\bf 1}_2, \\
  \tau_A\otimes\tau_2\otimes{\bf 1}_2, \\
  {\bf 1}_2\otimes \tau_A\otimes \tau_1,\\
  {\bf 1}_2\otimes \tau_A\otimes \tau_2,\\
  \tau_1\otimes{\bf 1}_2\otimes\tau_A,\\
  \tau_2\otimes{\bf 1}_2\otimes\tau_A,\\
  \tau_A\otimes\tau_A\otimes\tau_A.
\end{array}\label{c07}
\end{eqnarray}
The complex primitive maximal Clifford algebras $C(0,1)$ and $C(0,5)$ can be obtained from
$C(1,2)$ and $C(0,7)$, respectively, by deleting two gamma-matrices. {}From $C(0,7)$ we can,
e.g., consider the last tensor-product column, eliminate the two terms containing $\tau_1$ and
$\tau_2$ and replacing ${\bf 1}_2\mapsto 1$, $\tau_A\mapsto i$, to get
\begin{eqnarray}
C(0,5) &\equiv& \begin{array}{c}
  \tau_A\otimes\tau_1 , \\
  \tau_A\otimes\tau_2 , \\
  i \tau_1\otimes{\bf 1}_2,\\
  i \tau_2\otimes{\bf 1}_2,\\
  i \tau_A\otimes\tau_A.
\end{array}\label{c05}
\end{eqnarray}

It is worth pointing out that the $C(0,1)$ and $C(0,5)$ series were correctly considered as ``descendant"
series in \cite{crt1} due to the fact that they can be obtained from $C(1,2)$, $C(0,7)$ after
erasing extra-Gamma matrices. We find however convenient here to explicitly insert them in table
(\ref{bigtable}) and consider them as ``primitive", since they admit a different division-algebra
structure (they are almost complex, ${\bf C}$) w.r.t. the normal (${\bf R}$)-type maximal Clifford
algebras they are derived from.\par
The remaining primitive maximal Clifford algebras $C(0,x+8m)$, for positive integers $m=1,2,\ldots$ and
$x=1,3,5,7$, can be recovered from the $mod\quad 8$ properties of the Gamma-matrices. Let
$\overline{\tau}_i$ be a realization of $C(0,x)$ for $x=1,3,5,7$. By applying the (\ref{one}) algorithm  
to $C(0,7)$ we construct at first the
$16\times 16$ matrices realizing $C(1,8)$ (the matrix with
positive signature is denoted as $\gamma_9$, ${\gamma_9}^2 ={\bf
1}$, while the eight matrices with negative signatures are denoted
as $\gamma_j$, $j=1,2\ldots , 8$, with ${\gamma_j}^2 =-{\bf 1}$).
We are now in the position \cite{crt1} to explicitly construct the whole
series of primitive maximal Clifford algebras $C(0,x+8n)$,
through the formulas
\begin{eqnarray}
C(0,x+8n)&\equiv& \begin{array}{lcr} {\overline\tau}_i\otimes
\gamma_9\otimes \ldots&\ldots&\ldots\otimes\gamma_9,\\ {\bf
1}_4\otimes\gamma_j\otimes{\bf 1}_{16}\otimes\ldots & \ldots &
\ldots\otimes{\bf 1}_{16},\\
 {\bf 1}_4\otimes\gamma_9\otimes\gamma_j\otimes {\bf
1}_{16}\otimes\ldots &\ldots&\ldots\otimes{\bf 1}_{16},  \\ {\bf
1}_4\otimes\gamma_9\otimes\gamma_9\otimes\gamma_j\otimes {\bf
1}_{16}\otimes \ldots&\ldots&\ldots\otimes{\bf 1}_{16},  \\ \ldots
&\ldots&\ldots, \\ {\bf
1}_4\otimes\gamma_9\otimes\ldots&\ldots&\otimes
\gamma_9\otimes\gamma_j,
\end{array}\label{quatern}
\end{eqnarray}
Please notice that the tensor product of the $16$-dimensional representation is taken $n$ times.
 
\section{On division algebras.}
 
In the previous section we furnished a simple algorithm to explicitly construct any given
Clifford irrep of specified division-algebra type. It is convenient to review here the basic
features of division algebras which will be needed in the following.\par
The four division algebra of real (${\bf R}$) and complex (${\bf C}$) numbers, quaternions (${\bf H}$)
and octonions 
(${\bf O}$) possess respectively $0$, $1$, $3$ and $7$ imaginary elements $e_i$ satisfying the relations
\begin{eqnarray}
e_i\cdot e_j &=& -\delta_{ij} + C_{ijk} e_{k},
\label{octonrel}
\end{eqnarray}
($i,j,k$ are restricted to take the value $1$ in the complex case, $1,2,3$ in the quaternionic case and
$1,2,\ldots , 7$ in the octonionic case; furthermore, the sum over repeated indices is understood).\par
$C_{ijk}$ are the totally antisymmetric division-algebra structure constants. The octonionic division
algebra is the maximal, since quaternions, complex and real numbers can be obtained as its restriction.
The totally antisymmetric octonionic structure constants can be expressed as
\begin{eqnarray}
&C_{123}=C_{147}=C_{165}=C_{246}=C_{257}=C_{354}=C_{367}=1&
\end{eqnarray}
(and vanishing otherwise).
\par
The octonions are the only non-associative, however alternative (see \cite{bae}), division algebra.\par
Due to the antisymmetry of $C_{ijk}$, it is clear that we can realize (\ref{cliff}) by associating the
$(0,3)$ and $(0,7)$ signatures to, respectively, the imaginary quaternions and the imaginary octonions.\par
For our later purposes it is of particular importance the notion of division-algebra principal conjugation.
Any element $X$ in the given division algebra can be expressed through the sum
\begin{eqnarray}
X&=&x_0 + x_ie_i ,
\end{eqnarray}
where $x_0$ and $x_i$ are real, the summation over repeated indices is understood and the positive integral
$i$ are restricted up to $1$, $3$ and $7$ in the ${\bf C}$, ${\bf H}$ and ${\bf O}$ cases respectively.
The principal conjugate $X^\ast$ of $X$ is defined to be
\begin{eqnarray}\label{conjug}
X^\ast&=&x_0 - x_ie_i .
\end{eqnarray}
It allows introducing the division-algebra norm through the product $X^\ast X$. The 
normed-one restrictions $X^\ast X =1$ select the three parallelizable spheres $S^1$, $S^3$ and $S^7$
in association with ${\bf C}$, ${\bf H}$ and ${\bf O}$ respectively.
\par
Further comments on the division algebras and their relations with Clifford algebras can be found in \cite{crt1}
and \cite{bae}.   

\section{On fundamental spinors.}

In section {\bf 2} we discussed the properties of the Clifford irreps, presenting a method to explicitly
construct them and mentioning their division-algebra structure. It is worth reminding that the division-algebra character of fundamental spinors does not necessarily (depending on the given space-time) coincide with the division-algebra type of the corresponding Clifford irreps.\par
In the Introduction we already mentioned that the fundamental spinors carry a representation of the generalized Lorentz group with a minimal number of real components in association with the maximal, compatible, allowed division-algebra structure.\par
The following table, taken from the results in \cite{fer} and \cite{oku}, see also \cite{crt1}, presents the
comparison between division-algebra properties of Clifford irreps (${\bf\Gamma}$) and fundamental spinors (${\bf\Psi}$),
in different space-times parametrized by $\rho= s-t\quad mod\quad 8$.
We have

 { {{\begin{eqnarray}&\label{gammapsi}
\begin{tabular}{|c|c|c|}\hline
$\rho$&${\bf \Gamma}$ &${ \bf\Psi}$
\\ \hline

$0$&${\bf R}$&${\bf R}$\\ \hline

$1$&${\bf R} $&${\bf R}$\\ \hline

$2$&${\bf R}$&${\bf C}$\\ \hline

 $3$&${\bf C}
$&${\bf H}$\\ \hline

$4$&${\bf H}$&${\bf H}$\\ \hline

 $5$&${\bf H}
$&${\bf H}$\\ \hline

$6$&${\bf H}$&${\bf C}$\\ \hline

 $7$&${\bf C}
$&${\bf R}$\\ \hline

\end{tabular}&\end{eqnarray}}} }

It is clear from the above table that, for $\rho=2,3$, the fundamental spinors can accommodate a larger
division-algebra structure than the corresponding Clifford irreps. Conversely, for $\rho= 6,7$, the Clifford irreps
accommodate a larger division-algebra structure than the corresponding spinors.
In several cases this mismatch of division-algebra structures plays an important role. For instance in
\cite{lt1} a method was introduced to construct superconformal algebras based on the minimal division algebra
structure common to both Clifford irreps and fundamental spinors. This method can be straightforwardly modified to
produce extended superconformal algebras based on the largest division-algebra structure. The price to be paid,
in this case, would imply the introduction, for $\rho = 2,3$, of reducible Clifford representations and,
conversely, for $\rho=6,7$ of non-minimal spinors.\par
The reason behind the mismatch can be easily understood on the basis of the algorithmic construction of Section {\bf 2} and of table (\ref{bigtable}). Indeed, all the maximal, descendant Clifford algebras appearing in table
(\ref{bigtable}) have all block-antidiagonal Gamma matrices with the exception of a single Gamma matrix given by
$\left(\begin{array}{cc}
  {\bf 1} & 0 \\0 & -{\bf 1}
\end{array}\right)$. Therefore, all non-maximal Clifford algebras which are produced by erasing this extra Gamma matrix
(a detailed discussion can be found in \cite{crt1}) are of block-antidiagonal form. 
We recall now that the fundamental spinors carry a representation of the generalized Lorentz group whose generators
are given by the commutators among Gamma matrices, $\relax [ \Gamma_i, \Gamma_j]$. For the non-maximal Clifford algebras under considerations these commutators are all in $2\times 2$ block-diagonal forms, allowing to introduce a (generalized, in the sense specified in \cite{crt1}) Weyl projection for fundamental spinors, with non-vanishing upper or lower components.\par
It is convenient to explicitly discuss the simplest Minkowskian cases where the mismatch appears (the general procedure can
be straightforwardly read from table (\ref{bigtable})). 
In the ordinary $(3,1)$  space-time the (${\bf R}$) Clifford irrep is obtained as the non-maximal Clifford algebra
$(3,1)\subset (3,2)$, obtained from the maximal (${\bf R}$) $(3,2)$ after erasing a time-like Gamma matrix.
On the other hand, the fundamental complex spinors are obtained from the reducible Clifford representation
$(3,1)\subset (4,1)$, obtained by erasing a space-like Gamma matrix from the (${\bf C}$) Clifford irrep
$(4,1)$. \par 
In the other Minkowskian cases we get
\par
{\em i)} $(4,1)$: ${\bf \Gamma}$ coincides with the maximal Clifford $(4,1)$ (${\bf C}$), while 
${\bf \Psi}$ is constructed in terms of the reducible, non-maximal Clifford representation $(4,1)\subset (6,1)$ (${\bf H})$, \par 

{\em ii)} $(7,1)$: ${\bf \Gamma}$ coincides with the non-maximal Clifford $(7,1)\subset (7,2)$ (${\bf H}$), while 
${\bf \Psi}$ is constructed in terms of the reducible, non-maximal Clifford representation $(7,1)\subset (8,1)$ (${\bf C})$, \par 
{\em iii)} $(8,1)$: ${\bf \Gamma}$ coincides with the maximal Clifford $(8,1)$ (${\bf C}$), while 
${\bf \Psi}$ is constructed in terms of the reducible, non-maximal Clifford representation $(8,1)\subset (10,1)$ (${\bf R})$.

\section{On generalized supersymmetries.}

In this section we will discuss the generalized supersymmetries,
introducing at first the most physically important example, the minkowskian
eleven-dimensional $M$-algebra which, together with its $12$-dimensional 
$(10,2)$ counterpart, the so-called $F$-algebra, is based on real spinors
(Majorana and Majorana-Weyl respectively). Later, we will introduce generalized supersymmetries
for complex and quaternionic spinors.\par
At first we need to recall (see \cite{kt}) that three matrices, denoted as $A,B,C$, have to be introduced
in association with the three conjugations (hermitian, complex and transposition) acting on
Gamma matrices. Since only two of the above matrices are independent
we choose here, following \cite{crt1}, to work with $A$ and $C$. $A$ plays the role of the time-like $\Gamma^0$ matrix in
the Minkowskian space-time and is used to introduce barred spinors. $C$, on the other hand,
is the charge conjugation matrix. Up to an overall sign, in a generic $(s,t)$ space-time, $A$ and $C$
are given by the products of all the time-like and, respectively, all the symmetric (or antisymmetric) Gamma-matrices\footnote{Depending on the given space-time (see \cite{kt} and \cite{crt1}), there are at most two charge conjugations matrices,
$C_S$, $C_A$, given by the product of all symmetric and all antisymmetric gamma matrices, respectively. In
special space-time signatures they collapse into a single matrix $C$. This is true, in particular, for the maximal Clifford algebras entering the table (\ref{bigtable})}.
The properties of $A$ and $C$ immediately follow from their explicit construction, see \cite{kt} and \cite{crt1}.\par
In a representation of the Clifford algebra realized by matrices with real entries, the conjugation acts as the identity, see (\ref{conjug}). In this case the space-like gamma matrices are symmetric, while the time-like gamma matrices are antisymmmetric, so that $A$ can be identified with the charge conjugation matrix $C_A$.\par
For our purposes the importance of $A$ and the charge conjugation matrix $C$ lies on the fact that, in a 
$D$-dimensional space-time ($D=s+t$) spanned by $d\times d$ Gamma matrices, they allow to construct a basis for $d\times d$ (anti)hermitian and (anti)symmetric matrices, respectively. It is indeed easily proven that, in the real and the complex cases
(the quaternionic case is different and will be separately discussed later), the 
$\left( \begin{array}{c}
  D\\
  k
\end{array}\right)$ antisymmetrized products of $k$ Gamma matrices
$A{\Gamma}^{[\mu_1 \ldots \mu_k]}$ are all hermitian or all antihermitian, depending on the value 
of $k\leq D$. Similarly, the antisymmetrized products $C{\Gamma}^{[\mu_1 \ldots \mu_k]}$ are all 
symmetric or all antisymmetric.\par
For what concerns the $M$-algebra, the $32$-component real spinors of the $(10,1)$-spacetime
admit anticommutators
$\{Q_a,Q_b\}$ which are $32\times 32$ symmetric real matrices with, at most, $32+\frac{32\times 31}{2}= 528$
components. Expanding the r.h.s. in terms of the antisymmetrized product of Gamma matrices, we get that it can
be saturated by the so-called $M$-algebra
\begin{eqnarray}\label{Malg}
    \left\{ Q_a, Q_b \right\} & = & \left( A \Gamma_\mu \right)_{ab} P^\mu
      + \ \left( A \Gamma_{[\mu\nu]}\right)_{ab}
     Z^{[\mu\nu]} +
     \ \left( A \Gamma_{[\mu_1 \ldots \mu_5]}\right)_{ab}
     Z^{[\mu_1 \ldots \mu_5]}.
\end{eqnarray}
Indeed, the $k=1,2,5$ sectors of the r.h.s. furnish $11+55+462=528$ overall components. Besides the
translations $P^\mu$, in the r.h.s. the antisymmetric rank-$2$ and rank-$5$ abelian tensorial central charges,
$Z^{[\mu\nu]}$ and $Z^{[\mu_1\ldots \mu_5]}$ respectively, appear.\par
The (\ref{Malg}) saturated $M$-algebra admits a finite number of subalgebras which are consistent
with the Lorentz properties of the Minkowskian eleven dimensions. There are $6$ such subalgebras
which are recovered by setting either one or two among the three sets of tensorial central charges
$P^{\mu}$, $Z^{[\mu\nu]}$, $Z^{[\mu_1\ldots \mu_5]}$ identically equal to zero (a completely degenerate subalgebra is further obtained by setting the whole r.h.s. identically equal to zero).\par
The fact that the fundamental spinors in a $(10,2)$-spacetime are also 32-component, due to
the existence of the Weyl projection as discussed in Section {\bf  4},\footnote{The $(10,2)$ Clifford algebra is obtained from the maximal Clifford algebra $(11,2)$ appearing in table (\ref{bigtable}), with the
extra $\left( \begin{array}{cc}
  {\bf 1}&0\\
  0&-{\bf 1}
\end{array}\right)$ Gamma matrix employed to construct Weyl-projected spinors.} implies that the
saturated $M$-algebra admits a $(10,2)$ space-time presentation, the so-called $F$-algebra,
in terms of $(10,2)$ Majorana-Weyl spinors ${\tilde Q}_{\tilde a}$, ${\tilde a}=1,2,\ldots, 32$.\par
In the case of Weyl projected spinors the r.h.s. has to be reconstructed with the help of a projection operator which selects the upper left block in a $2\times 2$ block decomposition.
Specifically, if ${\cal M}$ is a matrix decomposed in $2\times 2$ blocks as
${\cal M} =\left( \begin{array}{cc}
  {\cal M}_1&{\cal M}_2\\
  {\cal M}_3&{\cal M}_4
\end{array}\right)$, we can define 
\begin{eqnarray}\label{pweyl}
P({\cal M}) &\equiv & {\cal M}_1.
\end{eqnarray} 
The saturated $M$-algebra (\ref{Malg}) can therefore be rewritten as
\begin{eqnarray}\label{Falg}
    \left\{ {\tilde Q}_{\tilde a}, {\tilde Q}_{\tilde b} \right\} & = & 
  P\left( {\tilde A} {\tilde\Gamma}_{{\tilde \mu}{\tilde \nu}}\right)_{{\tilde a}{\tilde b}}
   {\tilde Z}^{[{\tilde \mu}{\tilde \nu}]  }
      +  P\left( {\tilde A} 
      {\tilde \Gamma}_{[{\tilde \mu}_1 \ldots {\tilde \mu}_6]}\right)_{{\tilde a}{\tilde b}}
     {\tilde Z}^{[{\tilde\mu}_1 \ldots {\tilde \mu}_6]},
\end{eqnarray}
where all tilde's are referred to the corresponding $(10,2)$ quantities. 
The matrices in the r.h.s. are symmetric in the exchange ${\tilde a }\leftrightarrow {\tilde b}$.
This time the rank-$2$ and selfdual rank-$6$ antisymmetric abelian tensorial central charges,
${\tilde Z}^{[{\tilde \mu}{\tilde \nu}]  }$ and respectively ${\tilde Z}^{[{\tilde\mu}_1 \ldots {\tilde \mu}_6]}$,
appear. Their total number of components is $66+462=528$, therefore proving the saturation of the r.h.s. . The saturated equation (\ref{Falg}) is named the $F$-algebra.\par
For what concerns its consistent subalgebras which respect the $12$-dimensional Lorentz covariance,
besides the trivial one (i.e. a vanishing r.h.s.) only two of them are allowed. Either one can set
identically to zero the rank-$2$ 
(${\tilde Z}^{[{\tilde \mu}{\tilde \nu}]  }\equiv 0$)
or the self-dual rank-$6$ antisymmetric tensorial central charges
(${\tilde Z}^{[{\tilde\mu}_1 \ldots {\tilde \mu}_6]}\equiv 0$).\par
These two $11$-dimensional and $12$-dimensional examples were explicitly discussed for their 
physical relevance. On the other hand, the above construction can be straightforwardly repeated
for any given space-time admitting real fundamental spinors. If the real spinors $Q_a$ have
$n$ components, the most general supersymmetry algebra is represented by
\begin{eqnarray}\label{Mgen}
    \left\{ Q_a, Q_b \right\} & = & {\cal Z}_{ab},
\end{eqnarray}
where the matrix ${\cal Z} $ appearing in the r.h.s. is the most general $n\times n$
symmetric matrix with total number of $\frac{n(n+1)}{2}$ components. For any given space-time we
can easily compute its associated decomposition
of ${\cal Z}$ in terms of the antisymmetrized products of $k$-Gamma matrices, namely
\begin{eqnarray}
{\cal Z}_{ab} &=& \sum_k(A\Gamma_{[\mu_1\ldots\mu_k]})_{ab}Z^{[\mu_1\ldots \mu_k]},
\end{eqnarray}
where the values $k$ entering the sum in the r.h.s. are restricted by the symmetry requirement for the 
$a\leftrightarrow b$ exchange
and are specific for the given spacetime. The coefficients $Z^{[\mu_1\ldots \mu_k]}$ are the rank-$k$ abelian tensorial central charges.  \par
When the fundamental spinors are complex or quaternionic (in accordance with table (\ref{gammapsi}))
they can be organized in complex (for the ${\bf C}$ and ${\bf H}$ cases) and quaternionic 
(for the ${\bf H}$ case) multiplets, whose entries are respectively complex numbers or quaternions.\par
The real generalized supersymmetry algebra (\ref{Mgen}) can now be replaced by the most general complex or quaternionic supersymmetry algebras, given by the anticommutators among the fundamental
spinors $Q_a$ and their conjugate ${Q^\ast}_{\dot a}$ (where the conjugation refers to the principal
conjugation in the given division algebra, see (\ref{conjug})).
We have in this case
\begin{eqnarray}\label{Mhol}
    \left\{ Q_a, Q_b \right\} =  {\cal Z}_{ab}\quad &,& \quad \left\{ {Q^\ast}_{\dot a}, {Q^\ast}_{\dot b} \right\} =  {{\cal Z}^\ast}_{\dot{a}\dot{b}},
\end{eqnarray}
together with
\begin{eqnarray}\label{Mher}
\left\{ Q_a, {Q^\ast}_{\dot b} \right\} &=&  {\cal W}_{{a}\dot{b}},
\end{eqnarray}
where the matrix ${\cal Z}_{ab}$ (${{\cal Z}^\ast}_{\dot{a}\dot{b}}$ is its conjugate and does not contain new degrees of freedom) is symmetric,
while ${\cal W}_{{a}\dot{b}}$ is hermitian.\par
The maximal number of allowed components in the r.h.s. is given, for complex fundamental spinors
with $n$ complex components, by\\
{\em ia) } $n(n+1)$ (real) bosonic components entering the symmetric $n\times n$ complex matrix ${\cal Z}_{ab}$ 
plus\\
{\em iia)} $n^2$  (real) bosonic components entering the hermitian $n\times n$ complex matrix 
${\cal W}_{{a}\dot{b}}$.\par
Similarly, the maximal number of allowed components in the r.h.s. for quaternionic fundamental
spinors with $n$ quaternionic components is given by \\
{\em ib) } $2n(n+1)$ (real) bosonic components entering the symmetric $n\times n$ quaternionic matrix ${\cal Z}_{ab}$ 
plus\\
{\em iib)} $2n^2-n$  (real) bosonic components entering the hermitian $n\times n$ quaternionic matrix 
${\cal W}_{{a}\dot{b}}$.\par
The previous numbers do not necessarily mean that the corresponding generalized supersymmetry is indeed
saturated. This is in particular true in the quaternionic case, which will be discussed at length in Sections {\bf 6} and {\bf 7}.\par
Some further remarks are in order. We can expand the r.h.s. of (\ref{Mhol}) and (\ref{Mher}) in terms of the antisymmetrized product of Gamma matrices only when the division-algebra character of the Gamma matrices coincides with the division-algebra character of spinors. As discussed at length in Section {\bf 4}, for the spacetimes  presenting a spinor-versus-Clifford mismatch,
the supersymmetry algebra (\ref{Mhol}) and (\ref{Mher}) still makes sense by letting either non-fundamental spinors 
enter the l.h.s. 
or, conversely, reducible representations of the Clifford algebra enter the r.h.s. .\par 
It is convenient, in order to facilitate the
comparison among the different generalized supersymmetries (real, complex or quaternionic) that can be introduced in some given space-time, to summarize the above results in a table which specifies the maximal number of allowed (real) bosonic components for symmetric (${\bf SYM}$) and hermitian (${\bf HER}$) matrices (${\cal Z}$ and respectively ${\cal W}$)
in correspondence to real, complex or quaternionic spinors carrying the same number of (real) components. Obviously,
these spinors can only be employed in association with space-times which support the corresponding spinor representations. We have  
 { {{\begin{eqnarray}&\label{tabledimen}
\begin{tabular}{|c|c|c|}\hline
&${\bf SYM}$ &${ \bf HER}$
\\ \hline

${\bf R}(4)$&$10$&$$\\

${\bf C}(2)$&$6 $&$4$\\

${\bf H}(1)$&$4$&$1$\\ \hline

${\bf R}(8)$&$36$&$$\\

${\bf C}(4)$&$20 $&$16$\\

${\bf H}(2)$&$12$&$6$\\ \hline

${\bf R}(16)$&$136$&$$\\

${\bf C}(8)$&$72 $&$56$\\

${\bf H}(4)$&$40$&$28$\\ \hline

${\bf R}(32)$&$528$&$$\\

${\bf C}(16)$&$272 $&$256$\\

${\bf H}(8)$&$144$&$120$\\ \hline

${\bf R}(64)$&$2080$&$$\\

${\bf C}(32)$&$1056 $&$1024$\\

${\bf H}(16)$&$544$&$496$\\ \hline

\end{tabular}&\end{eqnarray}}} }
(please notice that the blanks associated to the ({\bf R}) spinors and the
${\bf HER}$ column are due to the fact that, in this case, the hermitian matrix
${\cal W}$ coincides with the symmetric matrix ${\cal Z}$). 
\par
It is evident from the above table (and in general from the counting of independent
(real) bosonic components given at the items {\em ia)} and {\em iia)} above) that
any real generalized supersymmetry admitting a complex structure can be re-expressed 
in a complex formalism with $n$-component complex spinors and total number of
$n(2n+1)$ (real) bosonic components split into $n(n+1)$ components entering the
symmetric matrix ${\cal Z}$ and $n^2$ components entering the hermitian matrix ${\cal W}$.
The situation is different in the quaternionic case. The quaternionic structure requires a restriction
on the total number of bosonic generators. $n$-component quaternionic spinors can be described as 
$4n$-component real spinors. However, the r.h.s. of a quaternionic (\ref{Mhol}) and (\ref{Mher})
superalgebra admits at most $4n^2+n$ bosonic components, instead of $8n^2+2n$ of the most general
supersymmetric real algebra. We will later show that the Lorentz-covariance further restricts the number
of bosonic generators in a quaternionic supersymmetry algebra.\par
We conclude this section mentioning the two big classes of subalgebras, respecting the Lorentz-covariance,
that can be obtained from (\ref{Mhol}) and (\ref{Mher}) in both the complex and quaternionic cases.
They are obtained by setting identically equal to zero either ${\cal Z}$ or ${\cal W}$, namely
\par
{\em I)} ${\cal Z}_{ab}\equiv   {{\cal Z}^\ast}_{\dot{a}\dot{b}}\equiv 0$, so that the only bosonic degrees
of freedom enter the hermitian matrix ${{\cal W}}_{{a}\dot{b}}$ or, conversely,
\par
{\em II)} ${{\cal W}}_{{a}\dot{b}}\equiv 0$, so that the only bosonic degrees of freedom enter 
${\cal Z}_{ab}$ and its conjugate matrix ${{\cal Z}^\ast}_{\dot{a}\dot{b}}$.\par
Accordingly, in the following we will refer to the (complex or quaternionic) generalized supersymmetries satisfying
the {\em I)} constraint as ``hermitian" (or ``type $I$") generalized supersymmetries, while the
(complex or quaternionic) generalized supersymmetries satisfying the {\em II)} constraint will be referred to
as ``holomorphic" (or ``type $II$") generalized supersymmetries.
It is worth mentioning that the hermitian-versus-holomorphic restriction of complex and quaternionic generalized
supersymmetries is not only mathematically meaningful, but potentially interesting for physical applications as
well. E.g., in \cite{lt3} it was proven that the analytical continuation of the (\ref{Malg}) $M$-algebra can be carried out to the Euclidean, the corresponding Euclidean algebra being a complex holomorphic supersymmetry. 
 
\section{On complex and quaternionic generalized supersymmetries.}

Let us summarize the previous section results. Generalized supersymmetries can be classified according to their division-algebra character ${\bf Y}$ (with ${\bf Y}\equiv {\bf R}, {\bf C}, {\bf H}$). They can be conveniently
labeled with a pair of division algebras as  ``${\bf X}{\bf Y}$", where ${\bf X}$ specifies whether spinors are realized as column vectors of real numbers (${\bf X}={\bf R}$), complex numbers (${\bf X}={\bf C}$) or
quaternions (${\bf X}={\bf H}$). Accordingly, generalized supersymmetries fall into different cases:\par
{\em i)} ${\bf RR}$,\par
{\em ii)} ${\bf RC}$ and ${\bf CC}$,\par
{\em iii)} ${\bf RH}$, ${\bf CH}$ and ${\bf HH}$.\par
In the ${\bf CC}$, ${\bf CH}$ and ${\bf HH}$ cases a suffix can be added, specifying whether we are dealing 
with a hermitian (type $I$, therefore ${\bf CC}_I$, ${\bf CH}_I$, ${\bf HH}_I$) or a holomorphic
(type $II$, ${\bf CC}_{II}$, ${\bf CH}_{II}$, ${\bf HH}_{II}$) generalized supersymmetry.
A closer inspection shows that the following identities hold for hermitian supersymmetries
\begin{eqnarray}\label{isom1}
{\bf RC} &\equiv & {\bf CC}_I
\end{eqnarray}
and
\begin{eqnarray}\label{isom2}
&{\bf RH}\equiv {\bf CH}_I\equiv {\bf HH}_I&.
\end{eqnarray}  
The first identity means that representing complex spinors in real notations is tantamount to realize a complex hermitian supersymmetry. The second set of identities holds for supersymmetries realized with quaternionic spinors.\par
It should be pointed out that the results furnished in the following, unless explicitly stated, are valid
for any complex or quaternionic supersymmetry, even the ones non-minimally realized (see table (\ref{gammapsi})),  
that is, the ones constructed either with reducible representations of the Clifford algebra or with non-minimal spinors.\par
In the following, for simplicity, it will be symbolically denoted as ``${\bf M}_k$" the space of $\left( \begin{array}{c}
  D  \\ k
\end{array}\right)$-component,
totally antisymmetric rank-$k$ tensors of a $D$-dimensional spacetime, associated to the
basis provided by the hermitian $A\Gamma^{[\mu_1\ldots\mu_k]}$ matrices (namely, entering ``type $I$" supersymmetries). Similarly, the rank-$k$ totally antisymmetric tensors associated to the symmetric
matrices $C\Gamma^{[\mu_1\ldots\mu_k]}$ and entering the type $II$, holomorphic, supersymmetries will be denoted
as ``${\cal M}_k$" (the symbol ``$M_k$" will be reserved to real, ``${\bf RY}$", supersymmetries).
\par
It is quite convenient to illustrate how complex and quaternionic supersymmetries work by discussing
specific examples. The extension of both reasonings and results to general spacetimes is 
in fact guaranteed by the algorithmic construction introduced in Section {\bf 2}.
Complex supersymmetries can be illustrated by the ${\bf C}$-type $(4,1)$ spacetime. Quaternionic supersymmetries,
on the other hand, by the ${\bf H}$-type $(5,0)$ Euclidean space.\par
The Clifford algebra of the ${\bf C}$-type spacetime $(4,1)$ is explicitly realized through the (${\bf C}$)
algorithmic lifting of table (\ref{bigtable}), given by $(0,1)\rightarrow (3,0)\rightarrow (4,1)$, while
the quaternionic $(5,0)$ space is obtained through the $(0,3)\rightarrow (5,0)$ lifting.
\par
Let us discuss the $(4,1)$-spacetime at first.  It should be noticed that, besides the complex lifting,
$(4,1)$ can be obtained from a real lifting (the first row in table (\ref{bigtable})), as a non-maximal Clifford algebra
contained in $(4,3)$ ($(4,1)\subset (4,3)$). We recall, following the language of \cite{crt1}, that maximal Clifford algebras are the ones directly obtained from the so-called ``primitive maximal Clifford algebras" through the
algorithmic lifting procedure of Section {\bf 2}, while non-maximal Clifford algebras are recovered from them
after deleting a certain number of Gamma matrices associated to the extra dimensions. The explicit procedure
for constructing non-maximal Clifford algebras has been  presented in \cite{crt1}.\par
The erasing of two time-like Gamma matrices from $(4,3)$ leading to $(4,1)$ can be done {\em without} respecting
the almost complex structure of $(4,1)$. One is therefore led to a representation of $(4,1)$ with
$8$ (real) components spinors (just like the ${\bf C}$-type representation), but endowed with an ${\bf R}$-structure,
instead of a ${\bf C}$-structure. The associated generalized supersymmetry would be of ${\bf RR}$-type, with a total number of $36$ expected bosonic components. Indeed, it can be easily checked that in $D=7$ (for the $(4,3)$ space-time) dimensions, the bosonic sector of the supersymmetry algebra is given by the
$1+35=36$ rank-$k$ tensors ${{M}_0}^{(D=7)}+{{M}_3}^{(D=7)}$. Expanding these tensors in the $D=5$-dimensional ($(4,1)$ spacetime) basis we are led to the following identifications
\begin{eqnarray} 
{{M}_0}^{(D=7)}+{{M}_3}^{(D=7)} &\equiv & {{M}_0}^{(D=5)}+{{M}_3}^{(D=5)}
+2\times {{M}_2}^{(D=5)}+{{M}_1}^{(D=5)},\label{d7d5}
\end{eqnarray}
where the counting of the components reads as follows
\begin{eqnarray}
1+35 &=& 1+ 10 + 2\times 10 +5.
\end{eqnarray}
The equation (\ref{d7d5}) above corresponds to the saturated bosonic sector of the ${\bf RR}$ generalized
supersymmetry in a $(4,1)$ spacetime.\par
Let us discuss now the two complex supersymmetries (${\bf CC}_I$ and ${\bf CC}_{II}$) associated with
the $(4,1)$ spacetime.\par
It can be easily shown that\par
{\em i)} in the ${\bf CC}_I$ case the bosonic sector is expressed as
\begin{eqnarray}
&{\bf M}_1+{\bf M}_3+{\bf M}_5&,
\end{eqnarray}
The expected $16$ bosonic components (real counting) of the saturated complex hermitian algebra are indeed recovered through
\begin{eqnarray}
16 &=& 5+10+1;
\end{eqnarray}
it should be noticed that the rank $k$ antisymmetric tensors {\em are not} related by the Hodge duality;
{\em ii)} in the ${\bf CC}_{II}$ case the bosonic sector is expressed as
\begin{eqnarray}
&{\cal M}_2+{\cal M}_3&,
\end{eqnarray}
whose total number of bosonic components, $10+10=20$, indeed saturates the number of bosonic components
for the complex holomorphic supersymmetry; in this case as well the rank-$2$ and rank-$3$ bosonic tensors 
{\em are not} related by Hodge duality (indeed one sector is real while the other one is completely imaginary
since the product of the five distinct gamma matrices is proportional to $i$). However, a reality constraint can
be {\em further imposed} on the bosonic sector of ${\bf CC}_{II}$. If this Lorentz-consistent constraint is applied, the total number of bosonic components corresponds to half the number of saturated bosonic components of the
complex holomorphic supersymmetry. This consistent reduction is a common feature of all complex holomorphic supersymmetries and not a special case of just the $(4,1)$ spacetime.\par
It should be noticed that the $36$ bosonic components of the saturated $(4,1)$ ${\bf RR}$ supersymmetry are recovered
from the $16+20$ bosonic components of the saturated hermitian and holomorphic supersymmetries. In a loose notation
we can symbolically write
\begin{eqnarray}
&{\bf RR} \approx {\bf CC}_I +{\bf CC}_{II}.&
\end{eqnarray}
The above decomposition is a common property of hermitian and holomorphic complex supersymmetries. \par
Let us now discuss the complex supersymmetries associated with a quaternionic spacetime such as $(5,0)$,
whose fundamental spinors are two-component quaternions. In this case we should expect that the total number
of bosonic components would not exceed $6$ for hermitian supersymmetry and respectively $12$ for holomorphic
supersymmetry (see table (\ref{tabledimen})). Performing the same analysis as before for $(5,0)$ we get\par
{\em i)} in the ${\bf CH}_I$ case the bosonic sector is expressed as
\begin{eqnarray}
&{\bf M}_0+{\bf M}_1&,
\end{eqnarray}
the expected $6$ bosonic components (real counting) of the saturated hermitian algebra are indeed recovered through
\begin{eqnarray}
6 &=& 1+5;
\end{eqnarray}
it should be noticed that now the rank $k$ antisymmetric tensors {\em are} related by the Hodge duality
(i.e. ${\bf M}_0\equiv {\bf M}_5$, ${\bf M}_I\equiv {\bf M}_4$);\par
{\em ii)} in the ${\bf CH}_{II}$ case the bosonic sector is expressed as
\begin{eqnarray}
&{\cal M}_2\equiv{\cal M}_3&,
\end{eqnarray}
the two sectos being related by Hodge-duality, so that the total number of bosonic components (real counting)
is given by $10$. In the ${\bf CH}_{II}$ case the reality condition discussed above for the ${\bf CC}_{II}$ case
is {\em automatically implemented}.\par 
Let us now discuss the quaternionic supersymmetries based on $(5,0)$.
For what concerns ${\bf CH}_I$, due to the (\ref{isom2}) isomorphisms (implying in particular
${\bf HH}_I\equiv {\bf CH}_I$),  we recover the same results as for the ${\bf CH}_I$ case above.
Quite different is the situation for the ${\bf HH}_{II}$ case. We recall that Clifford matrices are now expressed
with quaternionic entries and that the bosonic sector in the r.h.s. of the supersymmetry algebra is expressed
through symmetric matrices (related by transposition) with quaternionic entries. 
The problem with transposition applied to quaternions is that it does not respect the quaternionic composition
law. Indeed, if we try implementing transposition on imaginary quaternions we are in conflict with their
product since, e.g., $(e_1\cdot e_2)^T={e_2}^T\cdot {e_1}^T = - e_3 \neq {e_3}^T$ (the only consistent operation 
would correspond to setting ${e_i}^T=-{e_i}$, which in fact coincides with the principal conjugation already employed
in the construction of quaternionic hermitian matrices). In a $D$-dimensional quaternionic space-time, quaternionic Gamma matrices are split into $D-3$ matrices with real entries plus $3$ matrices given by the three imaginary quaternions $e_i$ multiplying a common
real matrix $T$. The totally antisymmetric products $(C\Gamma^{[\mu_1\ldots \mu_k]})_{ab}$ are no longer, for fixed
values of $k\geq 2$, all symmetric or all antisymmetric in the $a\leftrightarrow b$ exchange.
In the $(5,0)$ example that we are discussing, realized by $2\times 2$ quaternionic matrices, the basis of $12$ symmetric
and $4$ antisymmetric quaternionic matrices are recovered through the following decomposition.
The $12$ symmetric matrices are recovered from the $5$ symmetric matrices given by $C\Gamma^\mu$ plus $7$ symmetric matrices entering $C\Gamma^{[\mu\nu]}$. The three remaining matrices in $C\Gamma^{[\mu\nu]}$ are antisymmetric and, together with $C$, they provide the basis of the $4$ antisymmetric $2\times 2$ quaternionic matrices\footnote{Hodge duality holds in the ${\bf HH}_{II}$ case.}.
The Lorentz-covariance therefore requires that the ${\cal M}_2$ bosonic sector cannot be used in constructing
a quaternionic holomorphic supersymmetry for $(5,0)$. Bosonic generators can only enter
the $5$-dimensional holomorphic quaternionic supersymmetry in the ${\cal M}_1$ sector. This holomorphic quaternionic
supersymmetry admits at most $5$ bosonic generators. Since this number is less than $12$ entering the (\ref{tabledimen}) table,
the quaternionic holomorphic supersymmetry cannot be saturated. This is a common feature of all ${\bf HH}_{II} $
supersymmetries: they cannot possess bosonic tensorial central charges, but at most a single bosonic central charge
given by the ${\cal M}_0$ sector.
Moreover, for dimensions above $D=4$, they cannot be saturated. Their complete classification is provided
in the next section.\par
It is convenient to summarize this section results presenting a table with the basic properties of generalized
supersymmetries, valid in any given space-time signature supporting the corresponding supersymmetry.
In the table it is specified whether the Hodge duality holds and the total number of bosonic components of the
supersymmetry algebra. We have\footnote{Due to the (\ref{isom1}) and (\ref{isom2}) isomorphisms, some repetitions are found in the table below. However, we prefer to leave these repetitions for clarity reasons.} 
{ {{\begin{eqnarray}&\label{susiestab}
\begin{tabular}{|c|c|c|}\hline
SUSY&${Hodge}$ &${ d.o.f.}$
\\ \hline

${\bf RR}$&yes&$S$\\

${\bf RC}$&no&$S$\\

${\bf RH}$&yes&$S$\\

${\bf CC}_I$&no&$S$\\

${\bf CH}_I$&yes&$S$\\

${\bf HH}_I$&yes&$S$\\ 

${\bf CC}_{II}$&no&$S^\flat$\\

${\bf CH}_{II}$&yes&$S^\sharp$\\

${\bf HH}_{II}$&yes&$<$\\

 \hline

\end{tabular}&\end{eqnarray}}} }   

The third column specifies the total number of bosonic components (bosonic degrees of freedom).
``$S$" stands for a supersymmetry algebra with maximal number of saturated bosonic components.
``$S^\flat$" refers to the possibility of introducing a reality constraint such that the total number
of bosonic components is halved (w.r.t. the saturation number). ``$S^\sharp$" specifies in which cases 
the reality condition is automatically implemented, so that the total number of bosonic components
is half the number expected by saturation condition. Finally, ``$<$" specifies in which cases the saturation
of the bosonic components can not be reached due to the Lorentz-covariance requirement for the bosonic
sector (saturation only happens in $D=3,4$ dimensions). 

\section{Generalized supersymmetries of the quaternionic spacetimes.}

In this section we classify the complex and quaternionic generalized supersymmetries associated to
quaternionic space-times carrying quaternionic fundamental spinors (i.e. those spacetimes possessing the
quaternionic ${\bf H} $ entry in both the ${\bf \Gamma}$ and the ${\bf \Psi}$ columns of table (\ref{gammapsi})).\par
Up to $D=13$,
such spacetimes are given by the following list
{ {{\begin{eqnarray}&\label{qspacetimes}
\begin{tabular}{|c|c|c|}\hline

$D=3$&$(0,3)$\\ \hline

$D=4$&$(0,4)$, $(4,0)$\\ \hline

$D=5$&$(1,4)$, $(5,0)$\\ \hline

$D=6$&$(1,5)$, $(5,1)$\\ \hline

$D=7$&$(2,5)$, $(6,1)$\\ \hline

$D=8$&$(2,6)$, $(6,2)$\\ \hline

$D=9$&$(3,6)$, $(7,2)$\\ \hline

$D=10$&$(3,7)$, $(7,3)$\\ \hline

$D=11$&$(4,7)$, $(8,3)$, $(0,11)$\\ \hline

$D=12$&$(4,8)$, $(8,4)$, $(0,12)$, $(12,0)$\\ \hline

$D=13$&$(5,8)$, $(9,4)$, $(1,12)$, $(13,0)$\\ \hline

\end{tabular}&\end{eqnarray}}} }  

Due to the ${\bf CH}_I\equiv {\bf HH}_I$ isomorphism (see (\ref{isom2})), only three
different cases have to be considered, namely ${\bf HH}_I$, ${\bf CH}_{II}$ and ${\bf HH}_{II}$).

The following results do not depend on the signature of the space-time, but only on its dimensionality
$D$. The odd-dimensional $D$ spacetimes are maximal Clifford spacetimes (see Section {\bf 2}). They admit
a uniquely defined charge conjugation matrix. The even-dimensional spacetimes entering the (\ref{qspacetimes}) table above
are non-maximal. Their fundamental spinors are Weyl-projected (according to the discussion in Section {\bf 2}).
The associated generalized supersymmetry algebras are therefore constructed by requiring hermitian
(or, respectively, symmetric) conditions on the bosonic r.h.s., together with a {\em non-vanishing} upper-left
diagonal block (let's recall that the supersymmetry algebras involving Weyl-projected spinors require
the $P$ projection introduced in (\ref{pweyl})). These even-dimensional spacetimes support two inequivalent charge-conjugation
matrices $C_S$, $C_A$ (see the footnote at the beginning of Section {\bf 5}). They differ by a ``generalized $\Gamma^5$" matrix
$\Gamma^5=_{def}
\left( \begin{array}{cc}
  {\bf 1} & 0 \\
  0& -{\bf 1}
\end{array}\right)$ 
($C_A=C_S\Gamma^5$). However, since the projection onto the upper-left diagonal block
does not discriminate between $C_S$, $C_A$, it is irrelevant in the r.h.s. expansion for 
bosonic generators (see example the formula (\ref{Eufalg}) below for the Euclidean $F$-algebra) which charge
conjugation matrix is chosen.\par
Let us start with the hermitian quaternionic supersymmetry ${\bf HH}_I$. In association with each one of the
quaternionic spacetimes (\ref{qspacetimes}) the bosonic sector is decomposed in rank-$k$ antisymmetric tensors, with total number
of (real counting) bosonic components according to the table
{ {{\begin{eqnarray}&\label{hh1}
\begin{tabular}{|c|c|c|}\hline
spacetime&bosonic sectors&bosonic components\\ \hline
$D=3$&${\bf M}_0$& $1$\\ \hline

$D=4$&${\bf M}_0$& $1$\\ \hline

$D=5$&${\bf M}_0+{\bf M}_1$&$1+5=6$\\ \hline

$D=6$&${\bf M}_1$&$6$\\ \hline

$D=7$&${\bf M}_1+{\bf M}_2$& $7+21=28$\\ \hline

$D=8$&${\bf M}_2$&$28$\\ \hline

$D=9$&${\bf M}_2+{\bf M_3}$&$36+84=120$\\ \hline

$D=10$&${\bf M}_3$&$120$\\ \hline

$D=11$&${\bf M}_0+{\bf M}_3+{\bf M}_4$&$1+165+330=496$\\ \hline

$D=12$&${\bf M}_0+{\bf M}_4$&$1+495=496$\\ \hline

$D=13$&${\bf M}_0+{\bf M}_1+{\bf M}_4+{\bf M}_5$&$1+13+715+1287=2016$\\ \hline

\end{tabular}&\end{eqnarray}}} }  
Please notice from the above table that the hermitian quaternionic supersymmetry
saturates the bosonic sector, as expected.\par
In this one, as well as in the following tables, the signature of the space-times
does not affect the properties of the bosonic sector.\par
Let us now discuss the holomorphic supersymmetries associated with the
(\ref{qspacetimes}) quaternionic spacetimes. The complex holomorphic supersymmetry
${\bf CH}_{II}$ is characterized by the table
{ {{\begin{eqnarray}&\label{ch11}
\begin{tabular}{|c|c|c|}\hline
spacetime&bosonic sectors&bosonic components\\ \hline
$D=3$&${\cal M}_1$& $3$\\ \hline

$D=4$&${\widetilde{\cal M}}_2$& $3$\\ \hline

$D=5$&${{\cal M}}_2$&$10$\\ \hline

$D=6$&${\widetilde {\cal M}}_3$&$10$\\ \hline

$D=7$&${\cal M}_0+{\cal M}_3$&$1+35=36$\\ \hline

$D=8$&${\cal M}_0+{\widetilde{\cal M}}_4$& $1+35=36$\\ \hline

$D=9$&${\cal M}_0+{\cal M}_1+{\cal M}_4$&$1+9+126=136$\\ \hline

$D=10$&${\cal M}_1+{\widetilde{\cal M}}_5$&$10+126=136$\\ \hline

$D=11$&${\cal M}_1+{\cal M}_2+{\cal M}_5$&$11+55+462=528$\\ \hline

$D=12$&${\cal M}_2+{\widetilde{\cal M}}_6$&$66+462=528$\\ \hline

$D=13$&${\cal M}_2+{\cal M}_3+{\cal M}_6$&$78+286+1716=2080$\\ \hline

\end{tabular}&\end{eqnarray}}} }  

The tilde on the rank-$k$ (for $k=\frac{D}{2}$)
sectors ${\widetilde {\cal M}}_{\frac{D}{2}}$ specifies that they are self-dual
(as such, their total number of bosonic components, in the real counting, is given
by $\frac{1}{2}\left( \begin{array}{c}
  D\\
  \frac{D}{2}
\end{array}\right)$).\par
It should be noticed that the total counting of bosonic components in the third column implies
that the ${\bf CH}_{II}$ superalgebras admit (see table (\ref{tabledimen})) half the number of bosonic components expected for complex spinors of the corresponding size, in accordance with (\ref{susiestab}).
The recognition of this property becomes quite important when applied to the $D=11$ and $D=12$ rows
of the table above. Their total number of bosonic components ($528=\frac{1}{2}\times 1056$) 
coincides with the number of bosonic components entering the $M$-algebra (\ref{Malg}) and the $F$-algebra (\ref{Falg}).
In the next section we will see that their analytic continuation to the Euclidean is given by generalized supersymmetry algebras of ${\bf CH}_{II}$ type.\par 
The last two tables are devoted to the quaternionic holomorphic supersymmetries ${\bf HH}_{II}$.
According to the previous section discussion, we can state as a theorem that quaternionic holomorphic supersymmetries
{\em do not} involve bosonic tensorial central charges. The only admissible sectors are given by the following table
(together with the counting of their bosonic components)
{ {{\begin{eqnarray}&\label{hh11}
\begin{tabular}{|c|c|c|}\hline
spacetime&bosonic sectors&bosonic components\\ \hline
$D=3$&${\cal M}_0+{\cal M}_1$& $1+3=4$\\ \hline

$D=4$&${{\cal M}}_1$& $4$\\ \hline

$D=5$&${{\cal M}}_1$&$5$\\ \hline

$D=6$&$-$&$-$\\ \hline

$D=7$&$-$&$-$\\ \hline

$D=8$&$-$& $-$\\ \hline

$D=9$&${\cal M}_0$&$1$\\ \hline

$D=10$&${\cal M}_0+{{\cal M}}_1$&$1+10=11$\\ \hline

$D=11$&${\cal M}_0+{\cal M}_1$&$1+11=12$\\ \hline

$D=12$&${\cal M}_1$&$12$\\ \hline

$D=13$&${\cal M}_1$&$13$\\ \hline

\end{tabular}&\end{eqnarray}}} }  

The $mod\quad 8$ property of Clifford Gamma matrices allows to generalize the previous table
for quaternionic holomorphic ${\bf HH}_{II}$ supersymmetries in quaternionic space-times, according to 
{ {{\begin{eqnarray}&\label{holoquat}
\begin{tabular}{|c|ll|}\hline
$-$ &$D=0,6,7$&\quad mod \quad $8$\\ \hline

${{\cal M}}_0$& $D=1$&\quad mod \quad $8$\\ \hline

${{\cal M}}_1$&$D=4,5$&\quad mod \quad $8$\\ \hline

${\cal M}_0+{\cal M}_1$&$D=2,3$&\quad mod \quad $8$\\ \hline

\end{tabular}&\end{eqnarray}}} }  

The above results can be interpreted as follows. Quaternionic holomorphic ${\bf HH}_{II}$ supersymmetries
only arise in $D$-dimensional quaternionic space-times, where $D=2,3,4,5\quad mod \quad 8$.
No ${\bf HH}_{II}$ supersymmetry exists in $D=0,6,7\quad mod\quad 8$ $D$-dimensional spacetimes.\par In
$D=1 \quad mod\quad 8$ dimensions, ${\bf HH}_{II}$ supersymmetries only involve a single bosonic
charge. In this respect they fall into the class of quaternionic supersymmetric quantum mechanics, rather than
supersymmetric relativistic theories.\par
Finally, the ${\bf HH}_{II}$ supersymmetry algebra only admits a bosonic central charge in
$D$-dimensional quaternionic spacetimes for $D=2,3\quad mod\quad 8$.

\section{The $12$-dimensional Euclidean $F$-algebra.}

This section is devoted to furnish some physical motivations for our framework,
the notions that we have introduced and the results that have been obtained.
We will discuss some potentially interesting physical applications, especially in connection with the holomorphic
(${\bf CH}_{II}$) supersymmetry.\par
In \cite{lt3} the problem of finding an Euclidean counterpart to the $M$-algebra (\ref{Malg}) was addressed.
Euclidean $11$-dimensional spinors need to be complexified. For what concerns the bosonic generators,
on the other hand, their total number should not be changed in the Euclidean w.r.t. the Minkowskian spacetime.  
This means realizing an $11$-dimensional Euclidean supersymmetry with $528$ bosonic charges. By inspecting the table
(\ref{tabledimen}) and taking into account that $11$-dimensional spinors have $32$ complex components,
it becomes clear that such an Euclidean supersymmetry should be realized as a complex holomorphic supersymmetry
with extra reality condition (see table (\ref{susiestab})), which precisely means $\frac{1}{2}\times 1056 =528$ 
bosonic components.\par
The Euclidean $M$-algebra of reference \cite{lt3} is indeed realized as follows
\begin{eqnarray}\label{Eumalg}
\{ Q_a, Q_b\}&=& (C\Gamma^\mu)_{ab} P_\mu +(C\Gamma^{[\mu\nu]})_{ab} Z_{[\mu\nu]} +
(C\Gamma^{[\mu_1\ldots\mu_5]})_{ab} Z_{[\mu_1\ldots\mu_5]},
\end{eqnarray}
together with its complex conjugation. All remaining (anti)commutation relations are vanishing (the bosonic charges $P_\mu$, $Z_{[\mu\nu]}$, $Z_{[\mu_1\ldots \mu_5]}$ are abelian). In particular the relation 
$\{ Q_a, {Q_{\dot b}}^\ast\} =0$ is the holomorphic condition. The reality condition on the bosonic charges
provides the correct counting of the bosonic components. In \cite{lt3} it was further shown how to obtain
(\ref{Eumalg}) by analytic continuation of the Minkowskian $M$-algebra (\ref{Malg}).
We will prove now that, just like the Minkowskian $M$-algebra admits the $(10,2)$-spacetime $F$-algebra
(\ref{Falg}) counterpart, the Euclidean $M$-algebra (\ref{Eumalg}) admits a $12$-dimensional counterpart, also Euclidean,
which can be named ``the Euclidean $F$-algebra".\par
The construction goes as follows. The quaternionic space $(0,11)$ can be lifted, see (\ref{bigtable}), to the
quaternionic space $(13,0)$ ($(0,11)\rightarrow (13,0)$). The non-maximal Clifford $(12,0)\subset (13,0)$
dimensional reduction is a quaternionic space admitting Weyl-projected (in the sense specified in Section {\bf 2})
quaternionic fundamental spinors with total number, in real counting, of $64$ components.
Among all possible supersymmetries carried by such spinors, the complex holomorphic ${\bf CH}_{II}$ supersymmetry
(using the notation of Section {\bf 6}) is the only one which {\em automatically} provides a total number of
$528$ bosonic components.  The Euclidean $F$-algebra is given by the following relation
\begin{eqnarray}\label{Eufalg}
\{ {\widetilde Q}_a, {\widetilde Q}_b\}&=&  P({\widetilde C}{\widetilde \Gamma}^{[{\widetilde \mu}{\widetilde \nu}]})_{ab} {\widetilde Z}_{[{\widetilde \mu}{\widetilde \nu}]} +
P({\widetilde C}{\widetilde \Gamma}^{[{\widetilde \mu}_1\ldots{\widetilde\mu}_6]})_{ab} {\widetilde Z}_{[{\widetilde \mu}_1\ldots{\widetilde \mu}_6]},
\end{eqnarray}
(together with its complex conjugate, while all other (anti)commutators are vanishing). The tilde's refer to
quantities related to the $12$-dimensional Euclidean space $(12,0)$. The action of the operator $P$ on the r.h.s.
has been defined in (\ref{pweyl}).\par
The bosonic r.h.s. is given by a $66$-component rank-$2$ antisymmetric tensor plus a $462$-component
rank-$6$ selfdual antisymmetric tensor.
It should be pointed out that the dimensional reduction to $11$ dimensions immediately gives the Euclidean $M$-algebra
(\ref{Eumalg}). \par
It is likely that these complex holomorphic Euclidean algebras would play a role in formulating quantized
theories through functional integral. It is worth mentioning that, despite the fact that the $M$-theory itself has not yet been constructed,
in the literature, however, dynamical systems satisfying the supposed symmetry algebra of the
$M$-theory (see \cite{west}) have been introduced, see e.g. \cite{htz}
and references therein. A proper functional integral quantization of such systems could require the above Euclidean supersymmetries.
These considerations point towards a possible physical relevance of the mathematical construction discussed
in this paper.   

\section{Conclusions.}

This paper was devoted to perform a classification of (real, complex and quaternionic) generalized supersymmetries. We introduced the further notions of hermitian (complex and quaternionic) and holomorphic (complex and quaternionic) generalized supersymmetries and presented their classification in a series of tables, expressing their fundamental properties (such as the decomposition of their  bosonic sector, the total number of their bosonic components, the Hodge-duality related properties, etc.).\par
In order to be successful we clarified at first the necessary ingredients
entering such a classification (mainly the connection of Clifford algebras and spinors with division algebras).
The content of the paper is mainly mathematical. It should not be however dismissed the fact  that  its main motivation is based on physical considerations.
Indeed, within our classification we proved that the Euclidean analytic
continuation of the $M$-algebra of reference \cite{lt3} is an
$11$-dimensional complex holomorphic supersymmetry which also
admits a $12$-dimensional Euclidean presentation (see formula (\ref{Eufalg}), the so-called ``Euclidean $F$-algebra").  It should be stressed that these Euclidean counterparts of the Minkowskian $M$ algebra and of the $(10,2)$ $F$-algebra could find physical applications in performing the functional integral quantization of  dynamical systems admitting $M$-algebra related symmetries
(see e.g. \cite{west} and, for given examples of such systems, \cite{htz} and references therein).\par
The fact that the Euclidean $F$ and $M$ algebras arise as complex holomorphic algebras of ${\bf CH}_{II}$ type in our classification, clarifies why in \cite{lt3} the reality condition on the bosonic sector
(halving the number of bosonic degrees of freedom) had to be imposed.\par
We further discussed and classified quaternionic holomorphic supersymmetries. We did not attempt in this paper to find for them physical applications, but we succeeded in proving their
fundamental properties which can be stated as a theorem (see table
(\ref{holoquat})). Quaternionic holomorphic supersymmetries only exist in specific space-time dimensions. They {\em can not} possess higher rank bosonic tensorial central charges, but at most (depending on the dimensionality of the space-time) a single bosonic central charge.  \par
Finally, let us present a very partial list of physical topics and open problems where the present results
could hopefully find application (some of the topics here mentioned are actually under current investigation).
We have already mentioned in the text that superparticles models with tensorial central charges have been 
investigated in $4$-dimensional spacetimes (in \cite{rs} in terms of real spinors and in \cite{bl} in terms of complex spinors).
These models are linked with the Fronsdal proposal \cite{fro} of introducing bosonic tensorial coordinates for dealing with higher spin theories. This subject is related to the arising of a massless higher spin field theory in the tensionless limit of superstring theory and can be addressed within several different viewpoints (see e.g. \cite{vas}). For a recent review one can consult \cite{sor} and the references therein. 
Needless to say, our results pave the way for a unified description and generalization of the superparticle models with tensorial central charges (as well as of other dynamical systems presenting analogous features). \par
Another promising line of investigation concerns the deformation of superspace realized through the introduction of
non-anticommuting odd coordinates, assumed to satisfy a Clifford algebra (see \cite{ns}). This is presently an active branch of research. The mathematical results and the formalism discussed in this paper are quite suitable for investigating these deformations.\par
Let us finally comment that, despite the fact that in this paper only the ``generalized supertranslation algebras"
(in the loose sense discussed in the Introduction) were investigated and classified, the construction of
(extended) superPoincar\'e algebras can be straightforwardly recovered by adopting the procedure of \cite{lt1}. 
This construction requires that at first 
two separated copies of supertranslation algebras have to be introduced, so that the whole associated superconformal algebra can be extracted  by imposing the closure of the Jacobi identities. Next, the (extended) superPoincar\'e
algebra is obtained by means of a Inon\"u-Wigner type of contraction. 

$~$
\\$~$
\par
 {\large{\bf Acknowledgments.}} ~\\~\par

I have profited of precious discussions with H.L. Carrion, J. Lukierski, M. Rojas
and of helpful remarks by Z. Kuznetsova.

\end{document}